\documentclass{aastex}
\usepackage{emulateapj5}

\newcommand{\Fig}{Figure~}
\newcommand{\Tab}{Table~}
\newcommand{\Eq}{eq.~}

\slugcomment{Submitted to ApJ Letters on July 4 , 2001}
\revised{November 9, 2001}
\accepted{November 13, 2001}

\shorttitle{Iron emission line from BAL QSO}
\shortauthors{Oshima et al.}

\begin{document}

\title{Detection of an Iron Emission Feature from the Lensed BAL QSO
H1413+117 at $z=2.56$} \author{T.~Oshima\altaffilmark{1,2},
K.~Mitsuda\altaffilmark{1}, R.~Fujimoto\altaffilmark{1},
N.~Iyomoto\altaffilmark{1}, K.~Futamoto\altaffilmark{1},
M.~Hattori\altaffilmark{3}, N.~Ota\altaffilmark{1,4},
K.~Mori\altaffilmark{5}, Y.~Ikebe\altaffilmark{6},
J.M.~Miralles\altaffilmark{3,7,8}, and J-P.~Kneib\altaffilmark{9} }
\altaffiltext{1}{Institute of Space and Astronautical Science,
Sagamihara, Kanagawa, 229-8510, Japan}
\altaffiltext{2}{oshima@astro.isas.ac.jp}
\altaffiltext{3}{Astronomical Institute, T\^{o}hoku University,
Sendai, 980-8578, Japan}
\altaffiltext{4}{Department of Physics, Tokyo Metropolitan University,
Hachioji, Tokyo, 192-0397, Japan}
\altaffiltext{5}{Department of Earth and Space Science, Graduate
School of Science, Osaka University, Toyonaka, Osaka, 560-0043, Japan}
\altaffiltext{6}{Max-Planck-Institut f$\ddot{\rm u}$r
Extraterrestrische Physik, Postfach 1312, D-85741, Garching, Germany}
\altaffiltext{7}{ST-ECF, Karl-Schwarzschild Str. 2, Garching bei
M$\ddot{\rm u}$nchen D-85748, Germany}
\altaffiltext{8}{IAEF der Universit$\ddot{\rm a}$t Bonn, Auf
Dem H$\ddot{\rm u}$guel 71, D-53121 Bonn, Germany}
\altaffiltext{9}{Observatoire Midi-Pyr\'{e}n\'{e}es, UMR 5572, 14
Avenue E.Belin, F-31400 Toulouse, France}


\begin{abstract}
We present the X-ray energy spectrum of the lensed BAL QSO H1413+117
(the Cloverleaf) at $z=2.56$ observed with the {\sl Chandra} X-ray
observatory.  We detected 293 photons in a 40 ks Advanced CCD Imaging
Spectrometer (ACIS-S) observation.  The X-ray image consists of four
lensed image components, thus the photons are from the lensed QSO
itself.  The overall spectrum can be described with a power-law
function heavily absorbed by neutral matter at a redshift consistent
with the QSO redshift.  This supports the idea that intrinsic
absorption is significant for BAL QSOs.  The spectral fit
significantly (99 \% confidence) improves when we include an emission
line.  The centroid energy and intrinsic width (Gaussian $\sigma$) of
the line are $6.21 \pm 0.16$ keV and $220 ^{+270}_{-130}$ eV (90 \%
errors), respectively, in the QSO rest frame, assuming the absorbed
power-law as the continuum.  The equivalent width of the line in the
QSO rest frame is $960 ^{+1400}_{-480}$ eV.  We suggest that the large
equivalent width, the centroid energy, and the line broadness can be
explained by iron K emission arising from X-ray reprocessing in the
BAL flow, assuming it has a conical thin-sheet structure.
\end{abstract}


\section{Introduction}

H1413+117 (the Cloverleaf) is a QSO at a redshift of $z$=2.56,
gravitationally lensed into four image components with an angular
separation of $\sim$1$''$ \citep{Magain_etal_1988}; though the exact
nature of the lens object(s) remains elusive \citep{Kneib_etal_1998,
Kneib_Alloin_Pello_1998}.  The optical spectrum of the QSO shows broad
absorption features, hence it is classified as a broad absorption line
(BAL) QSO \citep{Hazard_etal_1984}.

BAL QSOs are in general extremely faint in X-rays
(Green et al. 1995; Green \& Mathur 1996, hereafter G1995; GM1996).
ROSAT, ASCA, and recent {\sl Chandra} studies have shown that BAL QSOs
are X-ray faint because of a large intrinsic absorption ($N_{\rm H} >
10^{23} {\rm cm}^{-2}$) and that the QSO itself is as bright as
typical radio quiet QSOs
(Gallagher et al. 1999; Brinkmann et al. 1999; Green et al. 2001,
hereafter G2001).  \citet{Wang_etal_2000} examined the ionization
state of the X-ray absorber and concluded that the absorbers for the
UV broad absorption lines and X-ray absorption can be the same,
although partial covering needs to be assumed for some of the UV
absorbers.  \citet{Elvis_2000} proposed an empirical model in which
narrow absorption line QSOs, BAL QSOs, and ordinary QSOs are unified.

X-ray emission from H1413+117 was detected at the 3 $\sigma$ level
with a 28 ks ROSAT PSPC observation \citep{Chartas_2000}.  We observed
H1413+117 with the {\sl Chandra} Advanced CCD Imaging Spectrometer
(ACIS-S) for 40 ks.  The high sensitivity of {\sl Chandra} and the
magnification of the gravitational lens enabled the detection of a
strong emission line from this distant BAL QSO.  In this letter we
report on the X-ray energy spectrum and discuss the implications for
BAL QSO X-ray emission : analysis of the X-ray image will be published
in a separate paper.

\section{Observation and results}

We observed H1413+117 on 2000 April 19 with the {\sl Chandra}
observatory.  The data were obtained using the back-illuminated
ACIS-S3 chip at a CCD temperature of $-120^{\circ}$C, and processed
using CIAO 2.1 with CALDB 2.6.
After the standard data screening, the net observation time was 38.2
ks.  The data contained a few background flares with about 2 times the
nominal background rate.  Since the contamination from the background
flares in the point source is small, we did not filter these periods,
however, we eliminated all the flares when making the background
spectrum.  A total of 293 X-ray events had spatial positions
consistent with H1413+117 within the systematic errors of {\sl
Chandra}'s attitude solution ($\sim$2$''$
\footnote{http://asc.harvard.edu/mta/ASPECT/celmon/index.htm}).  The
X-ray emission is spatially extended compared to the point spread
function of the X-ray mirror.  The extension is consistent with the
sum of the four gravitationally lensed images of the QSO.  The image
(\Fig\ref{fig:image}), reconstructed using the method proposed by
\cite{Tsunemi_etal_2001} and revised for faint sources by
\cite{Mori_etal_2001}, in which subpixel spatial resolutions can be
obtained for charge-split events, shows that X-ray photons come from
positions consistent with those of the four lensed images.  However,
the intensity ratios of the four images are not consistent with those
at optical wavelengths within the statistical errors; image A is
relatively brighter in X-rays than in the optical.  Further details of
the image analysis and their implications will be published in a
separate paper.

We then constructed an X-ray energy spectrum and performed model fits.
We created the telescope/detector response functions according to the
standard procedure provided by the {\sl Chandra} science center.  The
subtracted background spectrum was estimated from an annular region
with an outer radius of 130$''$
and an inner radius of 4$''$.  We confirmed that there is no
significant position dependency of the background counting rate by
changing the outer radius of the background region from 30$''$ to
130$''$.  The background counting rate is only 1\% of the source
counting rate, and is negligible compared to the statistical errors of
the source spectrum.  Before performing the model fit, we binned the
spectrum to a minimum of 20 photons per energy bin.

We first fitted the spectrum with a power-law function absorbed by
neutral matter.  We included two absorption components assuming the
solar abundance \citep{Anders_Grevesse_1989}; the absorption in our
Galaxy with a hydrogen column density $N_{\rm H}$ fixed to $1.78
\times 10^{20}~{\rm cm}^{-2}$ \citep{Stark_etal_1992}, and the
absorption at a redshift of $z_{\rm abs}$.  With $z_{\rm abs}$ set
free, it converged to $z_{\rm abs} = 2.55^{+0.27}_{-0.18}$ with
$N_{\rm H} = 2.4^{+1.5}_{-1.2} \times 10^{23} {\rm cm}^{-2}$ (90 \%
errors, see \Tab\ref{tab:spectral_fits}) and $\chi^2$=12.8 for 10
degrees of freedom (dof), well constrained by its iron K-edge.
However since the edge energy in the observer frame (2.00 keV) is
close to that of the Iridium M-V edge (2.04 keV) of the mirror
response function,
we need to be careful about the calibration.  We thus derived the
energy spectra of some QSOs in the archival data observed within 2
months of our observation.  We found that the spectra does not require
any additional absorption structure even if we vary the edge energy
around the Iridium edge.  We obtained $N_{\rm H} < 1\times 10^{22}
{\rm cm}^{-2}$ as the 90 \% upper limit of the absorption at $z_{\rm
abs}$=2.56 in the archival data, which is an order of magnitude
smaller than that we obtained for H1413+117.  Thus the absorber is
likely to be at the redshift of H1413+117 and we fix $z_{\rm abs}$ at
2.56 from now on.

Although the fit is acceptable at the 91 \% confidence level, the
residual of the fit shows
2-$\sigma$ excesses in adjacent bins around 1.7 keV ($\sim$6.2 keV in
the QSO rest frame).  We therefore masked out these bins
and those either side (\Fig \ref{fig:spec}(a)), and fitted an absorbed
power-law function.  The $\chi^2$ value became significantly smaller
($\chi^2$=2.7 for 7 dof).  The assumption that the residuals of those
two energy bins obey the same statistics as other energy bins is
rejected by an $F-$test at the 99 \% confidence level.  In
\Fig\ref{fig:spec}(a) we show the residuals of data against this
model, which indicates an emission line.  We thus tried including an
additional spectral component to describe the excess emission;
a Gaussian line feature and a power-law component of different
absorption column density.  In the latter model, we linked the
power-law index $\Gamma$ of the second power-law component with the
first one,
similar to the partial absorption model proposed by (G2001) if the
absorption column density of the second power-law component is small.
The best-fit $\chi^2$ value was 10.9 for 9 dof for the latter model
giving the improvement only at the 60 \% level significance.  On the
contrary, the resultant $\chi^2$ value of the Gaussian model was 2.8
for 8 dof and the improvement of the fit from the model without a line
was again 99 \% with an $F$-test.

In \Tab\ref{tab:spectral_fits}, we show the best-fit spectral
parameters.  The intrinsic absorption is again strong ($N_{\rm H} =
1.9^{+1.0}_{-0.8}\times 10^{23} {\rm cm}^{-2}$).  The centroid energy
$E_c$ of the line is $6.21\pm 0.16$ keV in the QSO rest frame.  Thus
the neutral iron K emission line at 6.4 keV is just outside the error
domain.  An intrinsically narrow emission line is also outside the 90
\% statistical errors (Gaussian $\sigma = 220^{+270}_{-130}$ eV).  The
equivalent width (EW) is $270 ^{+390}_{-130}$ eV ($960
^{+1400}_{-480}$ eV in the QSO rest frame).  We also performed a fit
forcing the line to be intrinsically narrow: in this case, the
improvement of the fit from a power-law is still significant at the
99\% level.  We obtained similar values for $E_c$ and EW.  We checked
the calibrations of the energy scale and the energy resolution of the
detector with archival data of SNRs and QSOs observed within 2 months
of our observation.  From the energy spectra accumulated within the
same image region in the detector coordinates that we used for
H1413+117, we found that the energy scale and the resolution are
correct within 2 \% ($\sim$120 eV) and 20 \% ($\sim$40 eV in FWHM),
respectively.  The {\sl Chandra} calibration web site
\footnote{http://asc.harvard.edu/ciao2.1/caveats/specres.html} states
that the uncertainty in the resolution is at most 40 \%.  Thus 6.4 keV
can be inside the error domain if we consider the calibration
uncertainty.  However the line still needs to be broad assuming an
absorbed power-law model as the continuum.  The line parameters also
depend on the choice of the continuum and because of the unusually
small $\Gamma=1.33^{+0.25}_{-0.48}$ we tested whether the addition of
a highly absorbed power-law, or a non-absorbed power-law component
gave any improvement.  The resultant fits for $\Gamma$ are $1.55\pm
0.6$, $1.34^{+0.7}_{-0.5}$ respectively, with larger statistical
errors.  Although the best-fit values of line parameters are not
sensitive to these choices of continuum, we obtained larger
statistical errors.  As a result, $E_c$=6.4 keV and zero intrinsic
width are included in the error domain.

 From the best-fit spectral model function we estimated the slope
$\alpha_{\rm OX}$ of a power-law connecting the rest frame 2500 $\AA$
and 2 keV flux densities (neglecting possible differences in the X-ray
and optical magnification factors) to be $3.31\pm 0.15$.  This is not
only steeper than the average $\alpha_{\rm OX}=1.57 \pm 0.15$ of
normal QSOs (G1995) but also steeper than that of the {\sl Chandra}
BAL survey QSOs (G2001).  Correcting for the intrinsic absorption we
obtain $\alpha_{\rm OX}=1.87\pm 0.15$ which is consistent with normal
QSOs, implying that BAL QSOs are obscured by strong absorption.
Finally, the X-ray luminosity in the 0.5--6 keV band (1.8--21 keV in
the QSO rest frame) is estimated to be $3.4 \times 10^{45}$ erg/s
assuming $H_0=$50 km/s/Mpc and $q_0 =0.5$, where we corrected only for
the galactic absorption.  The luminosity is $5.0 \times 10^{45}$ erg/s
if we also correct for the intrinsic absorption of the continuum.
Since the magnification of the lens is estimated to be $\sim$10
\citep{Chartas_2000}, the true luminosity of the QSO is a factor of 10
lower than the above values.

\section{Discussion}

The observed EW of the emission line, $960^{+1400}_{-480}$ eV in the
QSO rest frame, is an order of magnitude larger than those of typical
Seyfert~1 galaxies
(e.g. Mushotzky, Done, \& Pounds 1993).  The large EW suggests that a
considerable part of direct beam of the ionizing X-rays is blocked by
some intervening material, while the reprocessed emission is not
significantly absorbed.  Thus we need some low-ionization material
which partly covers the central engine and serves as an efficient
X-ray reprocessor.  A candidate for such beam blockers and/or
reprocessors is the BAL flow.  The BAL flow is believed to be a mass
flow with velocities of 1 -- 6 $\times 10^4$ km/s along the line of
sight.  If the flow is neither aligned with the accretion-disk plane
nor the disk axis, and if the system is axially symmetric, the flow
must have a conical structure, as proposed by \cite{Elvis_2000}.  The
BAL flow is believed to cover about 10--50 \% of the solid angle
around the central engine
(Krolik \& Voit 1998; Goodrich 1997).  Then the BAL flow may be a
conical thin sheet irradiated from the apex of the cone.  According to
\cite{Elvis_2000}, the conical BAL flow is optically-thick in the
direction of X-ray irradiation while it is optically thin in other
directions for photons above $^{>}_{\sim}$6 keV (see
\Fig\ref{fig:schematic}).  Then almost all X-ray photons from the
central source that enter the BAL flow are either absorbed or
scattered, while the most of re-emitted and scattered photons escape
from the flow \citep{Elvis_2000}.  Thus in spite of the small covering
solid angle, the conical BAL flow can be a more effective X-ray
reprocessor than a molecular torus or an accretion disk, for which the
reprocessing efficiency is limited by self-absorption.  Although a
conical BAL flow is not the only possibility, it is worth studying in
more detail.  Assuming the above geometry, the total intensities of
the iron K emission and the Thomson scattered photons are estimated
as,
\begin{equation}
F_{\rm FeK} \sim \eta \frac{E_{\rm K} F(E_{\rm K})}{\Gamma -1}
\frac{\Omega_{\rm BAL}}{4\pi}\omega_{\rm K} f_{\rm FeK},
\end{equation}
and
\begin{equation}
F_{\rm sca}(E) \sim F(E) \frac{\sigma_{\rm sca}}{\sigma_{\rm sca} +
\sigma_{\rm abs}(E)} \frac{\Omega_{\rm BAL}}{4\pi} ,
\label{eq:sca_spec}
\end{equation}
where $F(E)$ is the photon emission spectrum from the central X-ray
source integrated over $4\pi$ solid angle and is proportional to
$E^{-\Gamma}$, and $E_{\rm K}$, $\sigma_{\rm sca}$, $\sigma_{\rm
abs}(E)$, $\Omega_{\rm BAL}$, $\omega_{\rm K}$, and $f_{\rm FeK}$ are
respectively, the edge energy, the electron-scattering cross section,
the absorption cross section, the solid angle of the BAL flow, the
iron K fluorescence yield, and the fraction of iron K absorption over
the entire absorption for the photons above the K edge energy. The
factor, $\eta$, is defined as $\eta = \int_{E_{\rm
K}}^{\infty}(\sigma_{\rm abs}(E)/(\sigma_{\rm sca} + \sigma_{\rm
abs}(E)) F(E) dE/\int_{E_{\rm K}}^{\infty} F(E) dE$, and is 0.30 for
$\Gamma=1.8$, assuming the BAL flow to be neutral and of solar
abundance.

The highest EW, expected when the direct X-ray photon beam from the
central source is totally blocked, is
\begin{equation}
EW_{\rm max} = \frac{F_{\rm FeK}}{F_{\rm sca}(E_{\rm e})} 
             \sim 3.3~{\rm keV} \left(\frac{\eta}{0.3}\right)
              \left(\frac{f_{\rm FeK}}{1}\right)
              \left(\frac{\omega_{\rm K}}{0.35}\right),
\end{equation}
where $E_{\rm e}$ is the line energy.  On the other hand, when the
central engine is directly visible, the EW is the lowest;
\begin{eqnarray}
EW_{\rm min} &=& \frac{F_{\rm FeK}}{F(E_{\rm e})} \times \frac{1}{2}\nonumber\\
              &\sim& 60~{\rm eV}  \left(\frac{\eta}{0.3}\right)
              \left(\frac{\Omega_{\rm BAL}/4\pi}{0.15}\right)
              \left(\frac{f_{\rm FeK}}{1}\right)
              \left(\frac{\omega_{\rm K}}{0.35}\right).
\end{eqnarray}
The factor $1/2$ is included because one of the two BAL cones may be
behind the accretion disk and not be visible to the observer.  Thus
the large observed EW can be explained if most of the direct beam is
blocked by the BAL flow, which is generally believed to be the case
for BAL QSOs.  Then the observed continuum spectrum is dominated by
the scattered photons as described by \Eq\ref{eq:sca_spec}.  This
spectrum is better approximated by a partial absorption model (G2001)
than a single absorption model.  However in the present observation,
we do not have enough statistics to distinguish between these two.

Depending on the inclination angle to the line of sight and the cone
angle of the BAL flow, $\theta$, the iron emission line originating
from a certain portion of the BAL cone is either blue or red shifted.
For BAL QSOs, our line of sight is aligned with the BAL flow of a
certain azimuthal angle.  Then most of iron emission we observe is
from the far side of the cone, and is redshifted with energy shifts
ranging from 0 to $v_{BAL}/c \times | \cos(2\theta) |$ if $\theta>45$
deg.  Thus the center of the emission line is redshifted and the line
is broad.  For H1413+117, the velocity of the BAL flow is $\sim 0.04c
= 12000$ km/s \citep{Hazard_etal_1984} which implies a maximum energy
shift of $\sim$200 eV if $\theta \sim$70 deg.  Thus the energy shift
and the broadness of the line suggested from the present observation
are roughly consistent with these estimations.

Although the significance of the line itself is at the 99\% confidence
level, the line parameters are not well constrained because of the
limited statistics, possible calibration uncertainties, and also
possible uncertainties of the continuum spectrum.  Since the centroid
energy and the line broadening are key signatures of the reprocessing
in the BAL flow, further observations with better statistics and/or
better energy resolutions will be indispensable.

Finally we would like to point out that if the line of sight was
inside the BAL cone, we should observe a blueshifted emission line, or
both blueshifted and redshifted lines depending on the inclination
angle.  Since the direct beam would be visible in this case, the EW of
the line would be $\sim$60 eV.  These line features look similar to
the broad line features observed in some Seyfert galaxies which are
often described with disk-line models.  Thus reprocessing in the BAL
flow may at least partly account for those broad lines.  Recently the
detection of a redshifted broad strong (EW$\sim$600 eV) line feature
in the X-ray spectrum of a Seyfert~2 galaxy at $z \sim$1 is reported
(G. Hasinger 2001, private communication).  We conjecture that this
may have some connection to the emission line of H1413+117.

\acknowledgments
We thank P. Edwards for reviewing the manuscript.  K.M. is grateful to
W. Brinkmann for valuable discussions.  N.O. and N.I. are supported by
the JSPS Research Fellowship for Young Scientists.  JPK acknowledges
support from CNRS.

\clearpage

\begin{deluxetable}{ccccc}
\tablecaption{Results of Spectral Fits of the Total Spectrum.
\label{tab:spectral_fits}}
\tablewidth{0pt}
\tablehead{
\colhead{Parameter} & 
\colhead{Unit}   & 
\colhead{Power-law} &
\colhead{Power-law + line}   &
}
\startdata
$N~^{\rm a}$    & $10^{-5}~{\rm photon~keV}^{-1}~{\rm cm}^{-2} {\rm
s}^{-1}$ 
      & $24^{+25}_{-11}$ & $8.6^{+6.8}_{-4.2}$ \\
$\Gamma$   & 
      & $1.76\pm 0.50$ &  $1.33~^{+0.25}_{-0.48}$ \\
$N_{\rm H}~^{\rm b}$  & $10^{23}~{\rm cm}^{-2}$ 
   & 2.4 $^{+1.5}_{-1.2}$  & 1.9 $^{+1.0}_{-0.8}$  \\
$z_{abs}$ & &  $2.55^{+0.27}_{-0.18}$ & 2.56 (fixed) \\
$E_{e}~^{\rm c}$  &  keV  &  & $6.21 \pm 0.16$ \\
$\Delta E$ ($\sigma$)$~^{\rm c}$ & eV  &    & $220 ^{+270}_{-130}$ \\
$EW ~^{\rm c}$ & eV &     &  $960^{+1400}_{-480}$ \\
reduced $\chi^2$           & & 1.61 & 0.35  \\
d.o.f.                   & &10  & 8\\
\enddata


\tablenotetext{a}{The normalization of power-law functions. X-ray flux
at 1 keV in the QSO rest frame.}
\tablenotetext{b}{The hydrogen column density at the redshift of
$z_{abs}$.}
\tablenotetext{c}{The values in the QSO rest frame.}
\tablecomments{Galactic absorption of a fixed column density of $1.78
\times 10^{20}~{\rm cm}^2$ is included in both models.  The quoted
errors correspond to a single parameter error at 90\% confidence.}

\end{deluxetable}

\clearpage

\plotone{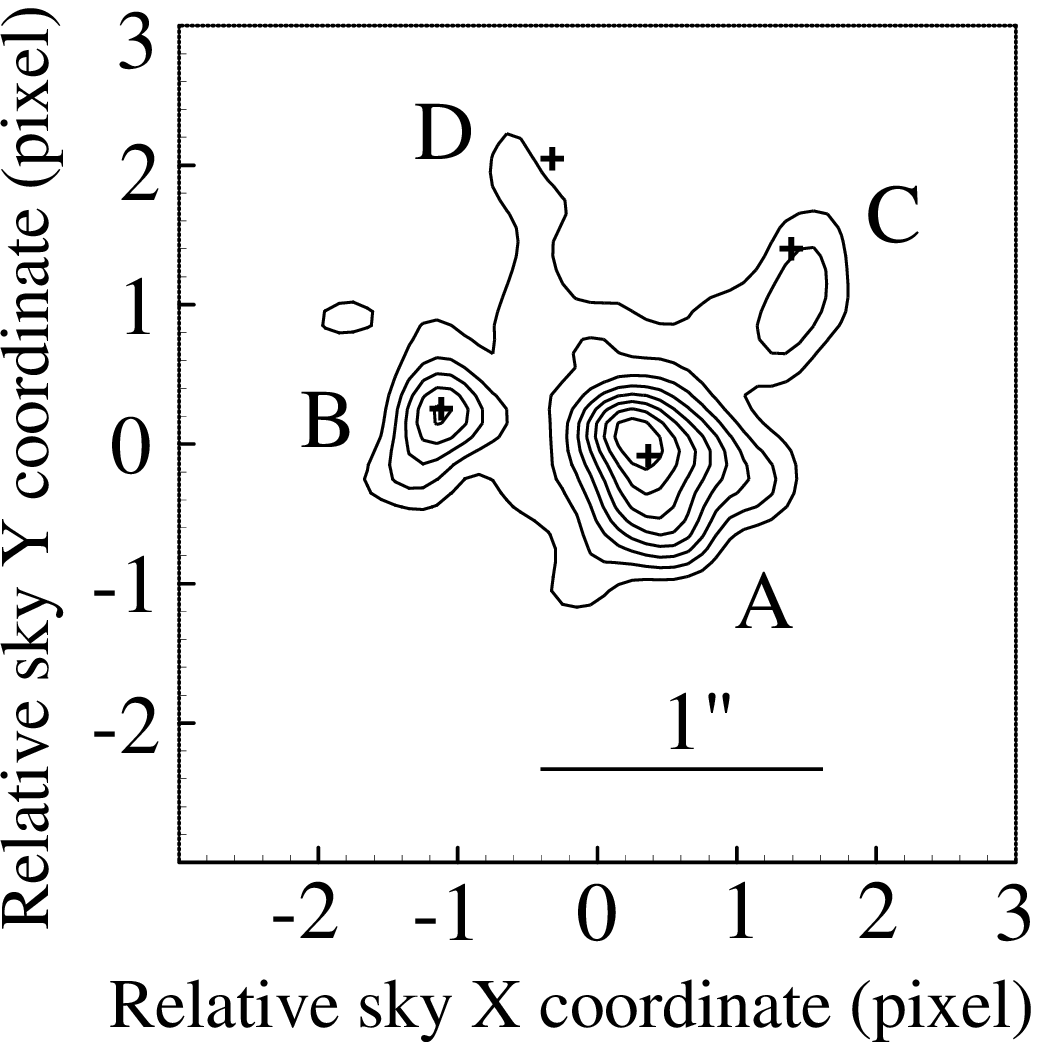}
\figcaption[f1.eps]{ACIS-S image of H1413+117.  Shifting the sky
coordinates of charge-split events according to the method proposed by
\citet{Tsunemi_etal_2001}, we obtained subpixel spatial resolution.
The image is then smoothed with a Gaussian function with a $\sigma$ of
0.2 pixel size. The X-ray surface brightness is shown with a lowest
contour level and an interval of respectively 17 and 8.5 counts/pixel.
The image consists of at least three point-like sources.  The relative
positions of the three sources are consistent with the multiply-lensed
images of the QSO A, B, and C at optical wavelengths, which are
shown with crosses for reference \citep{Chae_Turnshek_1999}.  We find
a hint of X-ray emission near the the optical position of image D.
\label{fig:image}}


\epsscale{0.9}
\plotone{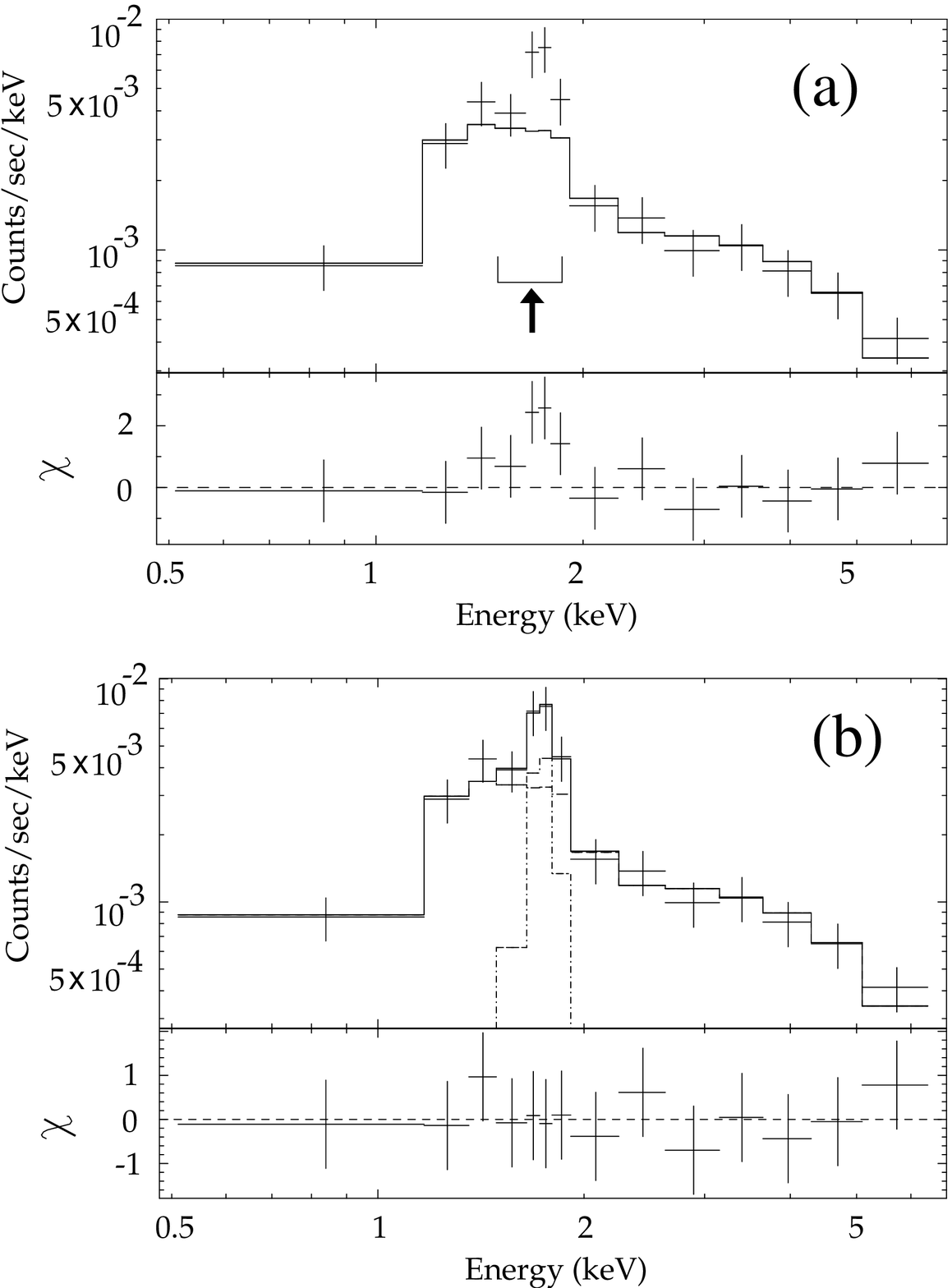}
\figcaption[f2.eps]{Spectral fittings of H1413+117. The observed
energy spectrum is shown with crosses and the model functions
convolved with the telescope and detector response functions are shown
with step functions. (a) An absorbed power-law model was assumed,
where the four energy bins marked with an arrow were not used.  (b) A
Gaussian model was added to the absorbed power-law model, using all
the bins.  The residual of the fits are shown in the bottom panels.
The vertical error bars are 1-$\sigma$ statistical errors.
\label{fig:spec}}

\clearpage

\epsscale{1}
\plotone{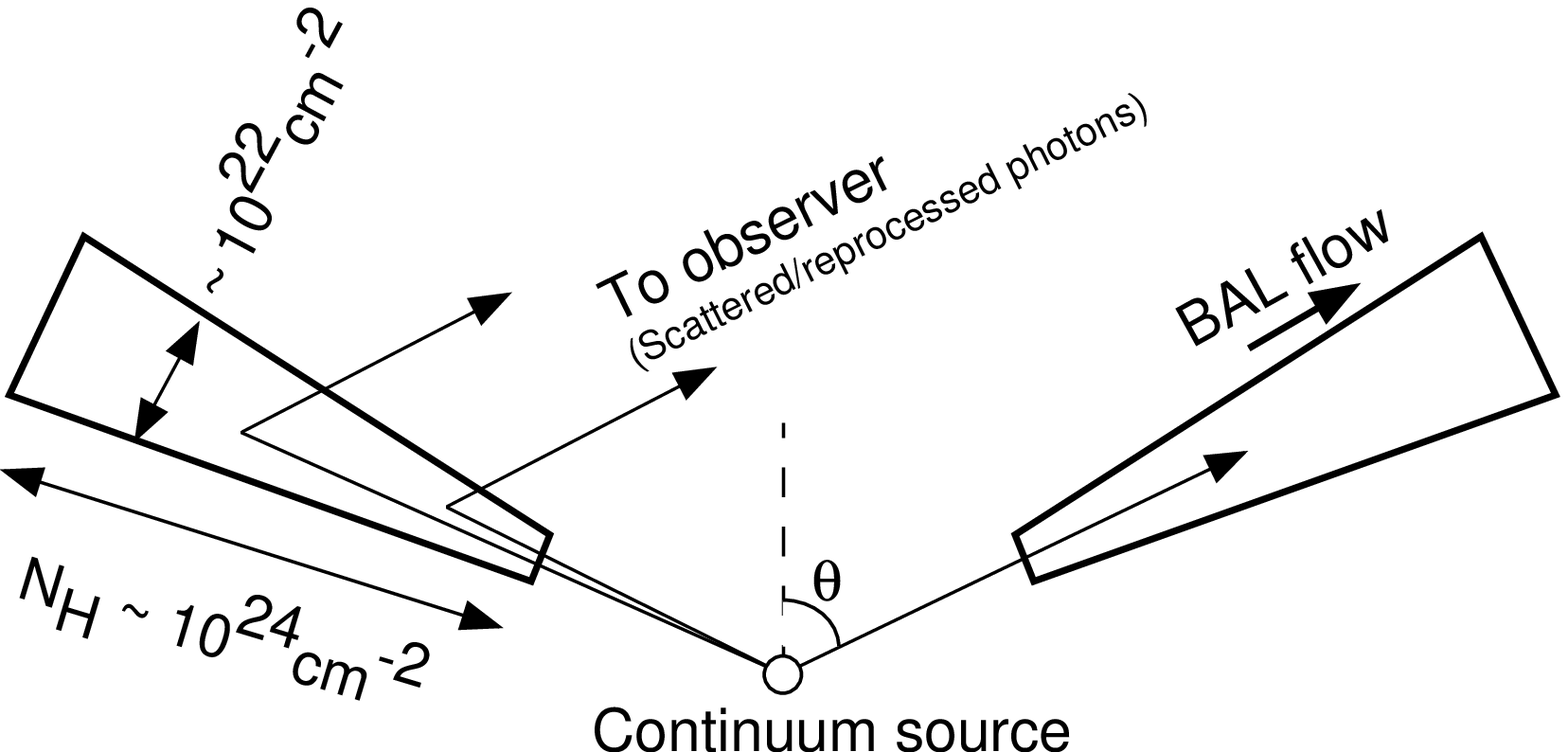}
\figcaption[f3.eps]{Schematic view of the conical BAL-flow model
according to the model of \citet{Elvis_2000}.  In this model, the flow
is optically thick for the X-ray photons from the central source, but
is optically thin for the scattered and the re-emitted photons.
\label{fig:schematic}}

\end{document}